# Robust, High-Contrast, Recyclable Zinc-Based Dynamic Windows via Synergistic Electrolyte and Interfacial Engineering

Fei Peng[1], Zhichao Liang[1], Zhoutao Yang[1], Yijia Chen[1], Wenjie Mai[1,2*], Chuanxi Zhao[1,2*]

[1] Siyuan Laboratory, Guangdong Provincial Engineering Technology Research Center of Vacuum Coating Technologies and New Energy Materials, Department of Physics, Jinan University, Guangzhou, Guangdong 510632, People's Republic of China.

[2] Guangdong Provincial Key Laboratory of Nanophotonic Manipulation, Jinan University, Guangzhou, Guangdong 511443, China

Email: wenjiemai@email.jnu.edu.cn, tcxzhao@email.jnu.edu.cn

**Abstract:**

Zinc-based electrochromic devices offer a sustainable route for dynamic optical management but are plagued by poor cycling stability due to irreversible zinc plating/stripping and side reactions. Herein, we report a robust, high-contrast, and recyclable zinc-based dynamic windows enabled by a synergistic electrolyte and interfacial engineering strategy. Molecular dynamics simulations and electrochemical analyses reveal a dual-ion cooperative mechanism that governs the reversibility: anions ($F^-$) with the strongest binding affinity guide uniform Zn deposition by stabilizing the inner solvation shell, while formate anions ($HCOO^-$) co-enriched at the interface facilitate smooth stripping via protonation during the oxidation process. This orchestrated interplay effectively eliminates "dead Zn" accumulation and dendrite growth. Consequently, the device demonstrates a record-high lifespan of 15,000 cycles with negligible degradation and maintains a large optical modulation of >50 %, along with multi-optical states (transparent, gray, black, and mirror), Furthermore, it achieves a large reflectance modulation (ΔR > 50 %) stable for over 2,000 cycles. This work establishes the recyclable zinc-based dynamic window as a scalable, high-performance alternative to conventional electrochromic systems, advancing the feasibility of solution-processed energy-saving windows in sustainable buildings.

# 1.Introduction

Buildings account for nearly 40 % of global energy consumption, with heating, ventilation, air conditioning (HVAC) systems being the primary contributors[1]. Indoor temperature control alone can constitute up to 50 % of a building's total energy load. Dynamic windows, which actively modulate solar heat and visible light transmission via an applied voltage, offer a compelling strategy to mitigate this energy burden. Unlike static low-emissivity (Low-e) windows, dynamic glazing can achieve energy saving of 10–20 % by reducing the reliance on HVAC system and artificial lighting[2]. While various technologies exist- such as suspended particle and liquid crystal devices exist, conventional electrochromic (EC) windows based on transition metal oxides ($WO_3$, $TiO_2$[2], and $MoO_3$[3]) have been the most extensively explored. Alternative EC materials, including polymers, nanoparticles, and redox-active molecules[4], have also been investigated. However, challenges such as high fabrication costs, limited scalability, and insufficient long-term cycling stability hinder their widespread practical implementation.

Reversible metal electrodeposition (RME) has recently emerged as a promising alternative paradigm. The operating principle of RME device (RMED) relies on the reversible electrochemical deposition of metal ions to modulate light absorption or reflection. Reversing the voltage bias dissolves the metal films, restoring transparency. Because metals can achieve near-total opacity at thicknesses of merely 20–30 nm, RME devices offer superior optical contrast and faster switching kinetics compared to their oxide-based counterparts[5]. Additionally, they offer benefits such as broad color palettes, low-voltage operation (<2 V), and low-power consumption. Various metals including Cu[4c, 5a], Ag[4d], Bi[4b], and Bi-Cu[5c]. While Ag-based devices exhibit excellent reflectivity across a wide spectral range, they are limited by high materials cost, slow response speeds, poor optical memory[6]. Similarly, Bi-Cu based RME device, despite their rich color states, as well as flexible compatibility[5b, 5c, 7], face challenges regarding cycling life. In this context, Zn has garnered significant attention due to its low cost, earth abundance, and the inherent colorlessness of $Zn^{2+}$ electrolyte. Previous studies have demonstrated that Zn-based RME devices can achieve faster coloration than Ag or Bi–Cu counterparts. Nevertheless, their cycling lifespan falls significantly short of ~50,000-cycle industry benchmark established by ASTM E2240-23 standards. This limitation stems primarily from the intrinsic instability of aqueous zinc electrolytes, which are prone to parasitic hydrogen evolution (HER, E= -0.76 V vs. *SHE*) and the formation of insulating passivation layers like $Zn(OH)_2$ or

ZnO layers, leading to rapid performance degradation[8].

To mitigate the issues associated with aqueous electrolytes, polar aprotic solvents such as dimethyl sulfoxide (DMSO) have been explored due to their wide electrochemical windows, which effectively suppress HER. Notably, Prof. Barile's group pioneered the design of DMSO electrolyte for Zn-based RMED, demonstrating promising long-term cycling and color-neutral switching[9]. Furthermore, while pure DMSO has a high freezing point (~18.5 °C), this can be mitigated by introducing cosolvents like acetonitrile, maintaining a weakly acidic electrolyte pH is also crucial to prevent $Zn(OH)_2$/ZnO precipitation[7, 9b]. Despite these advances, state-of-the-art Zn-RMEDs still achieve only ~10,000 stable cycles and require complex external voltage to kept EC performance[9b, 10]. Consequently, it remains a significant challenge to simultaneously achieve ultra-long-term durability, multi-stable optical states, high reflectivity, rapid switching kinetics in a single device.

In this work, we propose a comprehensive strategy involving synergistic electrolyte engineering and counter-electrode modification to develop advanced Zn-RMEDs. Through the rational design of the electrolyte composition- specifically the incorporation of halide anions, organic ligands, polymeric additives, alongside counter electrode interface engineering[11], we have fabricated a device that exhibits fast switching ($t_c$=5.5 s/$t_b$=7.1 s), and unprecedented cycling stability. The device exceeds 15,000 cycles for transmittance modulation and achieves over 2000 cycles for reflectance modulation ($\Delta R > 50$ %). Besides, our Zn-RMED maintains robust performance even under harsh conditions, demonstrating stable operation at 80 °C for over 3000 cycles. This work demonstrates a compelling platform that unites high performance and robust stability, advancing Zn-based systems as a viable candidate for the next-generation energy-saving windows.

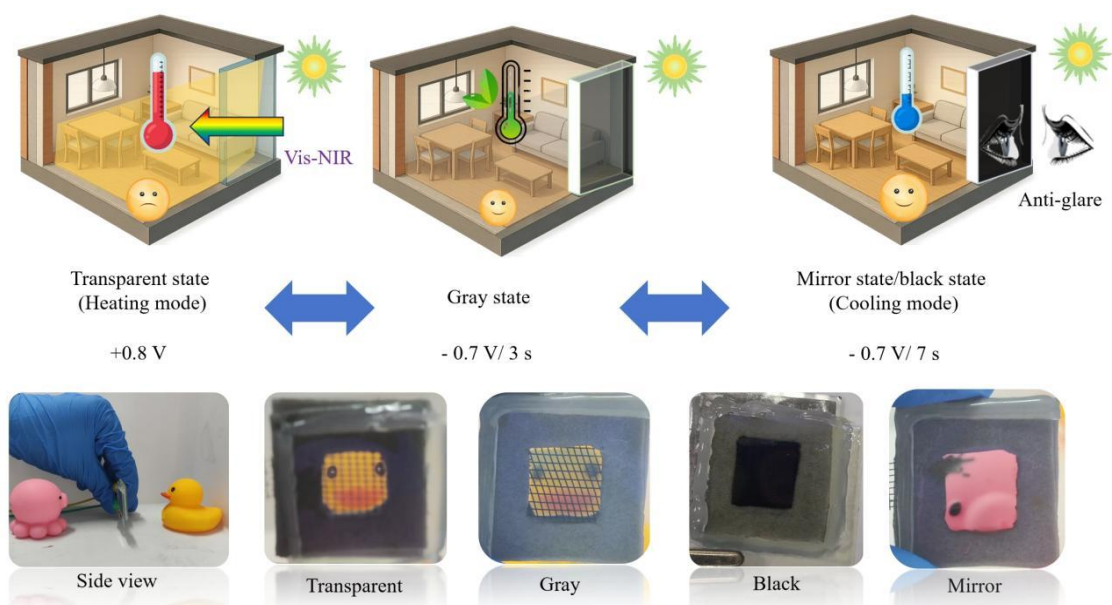

**Scheme 1. Working principle and multi-mode optical modulation of the Zn-RMED smart window.** (Top) Conceptual illustration of the energy-saving mechanism across 3 functional states: the Transparent state (passive heating mode, +0.8 V) maximizes solar transmission; the gray state (privacy mode, -0.7 V, 3 s) balances visible light transmittance; and the Mirror/Black state (cooling mode, -0.7 V, 7 s) effectively blocks solar heat gain (Vis-NIR). (Bottom) Photographic demonstration of the corresponding optical appearances.

## 2. Results and discussion

### 2.1. Operating Modes/Optical Modulation of the Zn-RMED

Selective modulation of solar radiation is a fundamental requirement for dynamic windows to accommodate seasonal temperature fluctuations. Specifically, the ability to actively manage near-infrared (NIR) radiation (heat) and visible light is critical for optimizing energy efficiency. Scheme 1 conceptually illustrates the working principle and the three distinct operating modes of the Zn-RMED. In mode I (transparent state), the Zn-RMED exhibits a transparent state, allowing maximum solar transmission to facilitate passive heating and natural lighting. While beneficial in winter, this high transmittance can lead to undesirable overheating during summer months if uncontrolled. Mode II (Gray State): Upon applying a voltage of -0.7 V for a short time of 3 s, the device switches to Mode II (gray state). In this state, a thin zinc film is deposited on the ITO electrode. This layer reflects partial solar radiation, effectively balancing indoor thermal comfort with privacy protection, while still maintaining a degree of visual connection to the outside. Mode III (Mirror/Block State): Prolonging the deposition time to 10 s at the same potential induces the formation of a dense, continuous Zn mirror (Mode III). This corresponds to the "black" or "mirror" state, where the optical transmittance is significantly suppressed to <2 %. This mode offers maximum solar blocking capability, making it advantageous for minimizing cooling loads and enhancing building energy efficiency in hot climates. The accompanying photographs (bottom panel) visually demonstrate the optical performance in each mode. Background objects - a yellow duck representing an indoor observer and a pink octopus representing an outdoor observer - are used to highlight the distinct visibility and privacy levels achieved across the transparent, gray, and mirror states.

**2.2. Electrolyte Design and Operating Principle of the Zn-RMED**

The architecture of the Zn-RME device and the chemical components of the electrolyte are depicted in **Figure 1a**. The device employs a sandwich configuration, comprising an ITO-coated glass working electrode, the electrochromic gel electrolyte, and a glucose-modified Zn mesh as counter electrode. As shown in Figure 1b, core innovation lies in the rationally designed electrolyte system, composed of $ZnBr_2$, lithium formate (LiCOOH), and lithium fluoride (LiF) dissolved in anhydrous DMSO with PVA as a gelling agent. Briefly, the fabrication involved injecting the gel electrolyte into the ITO-tank, then separated ITO (working electrode) from the Zn electrode by a spacer (3M tape) and sealing by hot melt adhesive. Figure 1c schematically illustrates the operating principle of the Zn-RMED, which relies on reversible electrochemical deposition and dissolution of Zinc between working and counter electrodes. Under a reductive bias of –0.7 V (Figure 1c(i)), $Zn^{2+}$ ions migrate to the ITO electrode and are reduced to form a metallic zinc film. This film effectively blocks and reflects incident light, switching the device to a "cooling mode" (black or mirror state). Conversely, applying a positive bias oxidizes the zinc film, stripping the metal back into the electrolyte and restoring the "heating mode" (transparent state) (Figure 1c(ii)). Notably, the use of a Zn mesh counter electrode facilitates the dynamic redistribution of $Zn^{2+}$, mitigating concentration polarization and maintaining ionic homogeneity to enhance cycling stability.

To decipher the atomistic origin of the high reversibility, molecular dynamics (MD) simulations were conducted to analyze the solvation structure and interfacial interactions. The radial distribution function (RDF, Figure 1d) reveals that DMSO molecules dominate the primary solvation shell (r~0.5 nm) due to their high concentration, forming a stable solvation sheath that ensures sufficient ionic mobility. However, interaction energy analysis (Figure 1f) uncovers a critical competitive coordination mechanism. Specifically, the F$^-$ anion exhibits the strongest binding affinity with $Zn^{2+}$(~-200 kJ mol$^{-1}$), significantly higher than that of DMSO. This strong interaction enables F$^-$ to displace solvent molecules at the electrode interface, guiding uniform Zn nucleation and facilitating the formation of a stable, fluoride-rich solid electrolyte interphase (SEI) that suppresses side reactions.

Crucially, the simulation also highlights the role of formate (HCOO$^-$) and PVA. HCOO- exhibits the second-strongest interaction (~-170 kJ mol$^{-1}$), leading to its co-enrichment at the interface. During the oxidative stripping process, the protonation of these interfacial formate anions (forming formic acid species) actively promotes the dissolution of the Zn film, thereby preventing the accumulation of "dead Zn." Meanwhile, the polymer additive PVA, despite a weaker interaction energy (~ -20 kJ mol-1), plays a vital role as a steric regulator. By selectively adsorbing onto active growth sites via long-range interactions, PVA homogenizes the ion flux and physically suppresses dendritic protrusion. Experimental validation via Scanning Electron Microscopy (SEM) confirms this synergistic mechanism (Figure 1e). Electrolytes lacking LiCOOH result in large, coarse particles that are difficult to strip completely, while those without LiF yield poor coverage and uniformity. In sharp contrast, the optimized electrolyte containing both anions and PVA yields a smooth, compact, and dendrite-free morphology. This validates that the "dual-ion synergy"—where F$^-$ anchors stable deposition and HCOO- facilitates clean stripping—combined with

PVA's steric regulation, is the fundamental driver of the device's record-breaking durability.

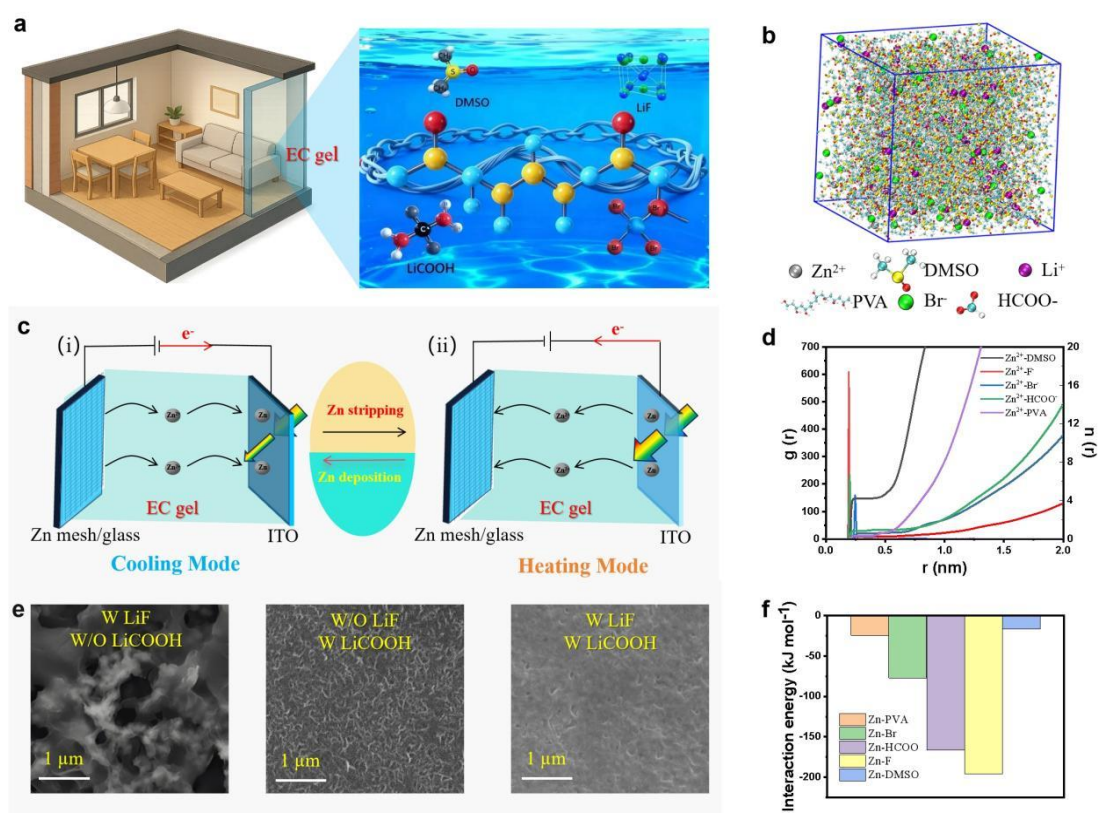

**Figure 1 Design, operation, and mechanism of the Zn-RMED**. (a) Device structure integrated into a smart building scenario with key electrolyte components. (b) MD simulation snapshot of the multi-component electrolyte system. (c) Working principle showing reversible Zn deposition (Cooling Mode, -0.7 V) and stripping (Heating Mode).(d) Radial distribution functions and coordination numbers of $Zn^{2+}$, revealing a DMSO-dominated primary solvation shell.(e) SEM morphology comparison of Zn deposits: without LiCOOH (left), without LiF (middle), and optimized synergistic electrolyte (right).(f) Interaction energy analysis highlighting the strong affinity of F- (anchoring deposition) and HCOO- (facilitating stripping).

**2.3. Optimization of Electrolyte Composition and Synergistic Mechanism**

To identify the optimal electrolyte architecture, we systematically screened alternatives by varying cations and anions, replacing LiCOOH and LiF with LiBr, LiCF$_3$COO, CH$_3$COOLi, Zn(CH$_3$COO)$_2$, and NaCH$_3$COO (Table 1). In Figure S1a, the control group based on ZnBr$_2$/Zn(CH$_3$COO)$_2$/NaCH$_3$COO/H$_3$BO$_3$/PVA formulation) afforded limited optical modulation ($\Delta T$~20 %). In contrast, while the ZnBr$_2$/LiBr/CH$_3$COOLi/PVA system (Figure S1b) achieved a deep colored state (T$_{min}$ ~1 %), it suffered from sluggish coloration kinetics ( >30 s). Similarly, electrolytes containing LiCF$_3$COO (Figures S1c, S1d) achieved excellent blocking capabilities ($\Delta T$ <0.1 %) yet were suffered from slow switching speeds (>20 s) and the limited cycling life (~800 cycles). The ZnBr$_2$/LiBr/LiF/PVA system (Figure S1e) also exhibited suboptimal coloration kinetics and ΔT. By comparison, our optimized electrolyte enables rapid, reversible, and durable switching, underscoring the efficacy of rational electrolyte engineering. **Figure 2a** compares the optical modulation performance among of 7 types of electrolytes, highlighting that our optimized device achieves the shortest deposition and stripping times among all tested electrolytes. The ionic conductivities and key performance metrics (coloration/bleaching time, optical reversibility, and cycle life) are summarized in Figure 2b and Table S1.

The superior performance stems from the critical, synergistic functions of each electrolyte component. (1) ZnBr$_2$ serves as the as the primary source of active Zn$^{2+}$ ions, while Br$^-$ are hypothesized to form [ZnBr$_x$]$^{2-x}$ complexes. This complexation modulates the deposition kinetics and suppresses dendrite growth[4a]. (2) Lithium formate (LiCOOH) serves a dual role. Li$^+$ enhances the ionic conductivity, while the formate ion (HCOO⁻) creates a weak acidic buffering environment (via the dissociation of its conjugate acid, pKa=3.75), which favors its dissociation tendency in aqueous solution and naturally establishes a weakly acidic environment that mitigates the formation of undesirable hydroxide or oxide by-products[8b]. Crucially, as indicated by our simulations, interfacial formate promotes smooth stripping via proton-assisted dissolution mechanisms. (3) Inspired by interfacial engineering in lithium-ion batteries, LiF is incorporated to promote the in-situ formation of a robust interface. The presence of F⁻ facilitates the in-situ formation of a $ZnF_2$ on the electrode surface. Due to the high electronegativity of fluorine, F⁻ weakens the interaction between $Zn^{2+}$ and solvent molecules[12], thereby reducing the desolvation energy barrier and accelerating deposition kinetics[13]. Finally, the PVA polymer serves as a rheological modifier, increasing electrolyte viscosity to homogenizes the ion diffusion field, leading to more compact dendrite-free zinc deposits and enhancing light-blocking efficiency.

Figure 2c demonstrates that the addition of LiF on deposition significantly accelerates deposition kinetics in the range of -0.5 V to -0.8 V, resulting in lower transmittance compared to the LiF-free counterpart within the same time. Conversely, Figure 2d confirms that faster LiCOOH plays a dominant role in accelerating the stripping process, aligning with the "stripping-facilitator" mechanism. The structural impact is further evidenced by XRD analysis (Figure 2e). The film deposited from the optimized electrolyte (containing both additives) exhibits a highly preferred (002) orientation, with an $I_{(002)}/I_{(101)}$ ratio reaching 24.73. This crystallographic texture corresponds to a denser, more uniform Zn film, which directly translates to higher device reflectivity. A comprehensive assessment is provided by the radar chart in Figure 2f, which clearly illustrates that the optimized electrolyte employed in this work outperforms other formulations in terms of rate capability (deposition/stripping) and cycling stability. Detailed EC performance metrics are listed in Table S1. Electrochemical impedance spectroscopy (EIS) data for the various electrolytes are presented in Figure S2.

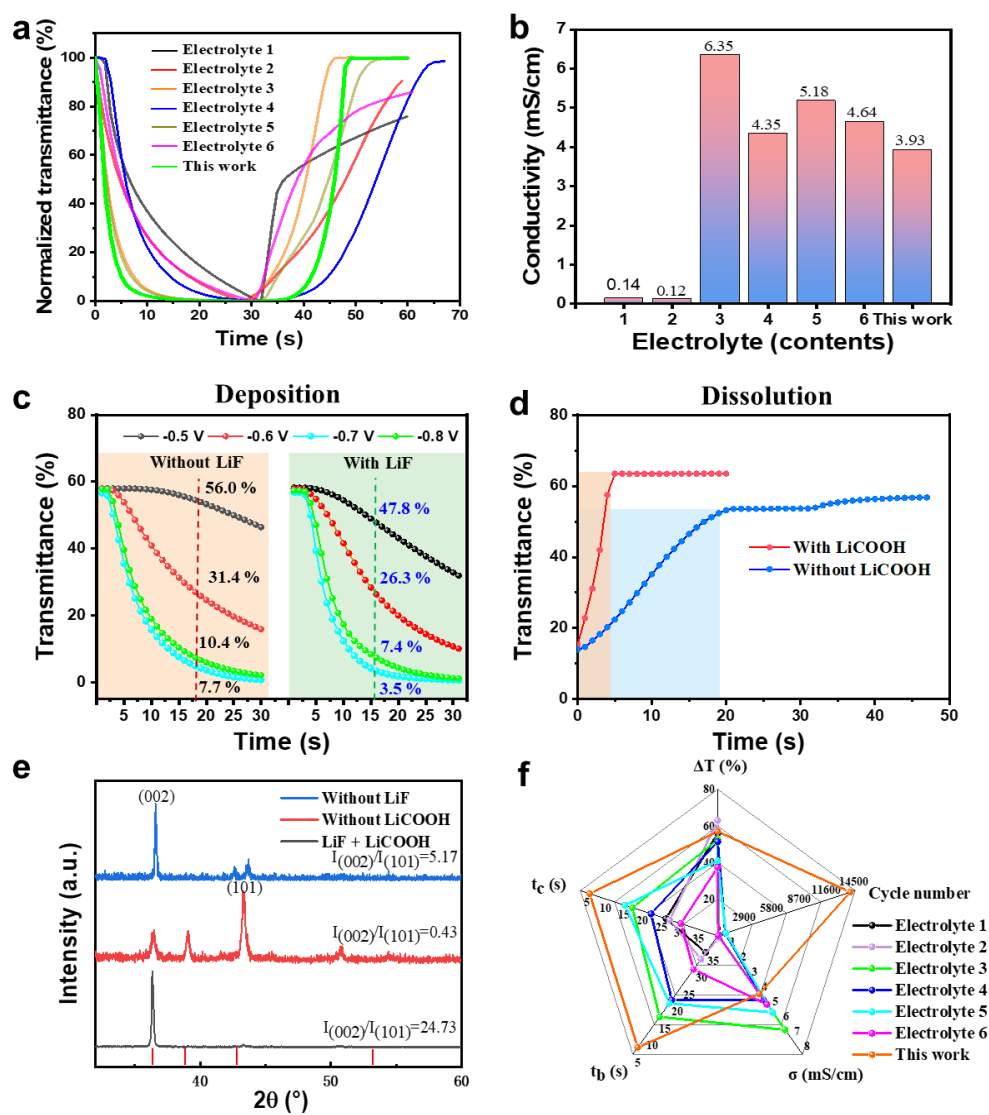

**Figure 2 Electrolyte screening and experimental validation of synergistic effects.** (a) Normalized deposition and stripping times comparing the optimized electrolyte

with control formulations. (b) Ionic conductivity comparison of the investigated electrolytes. (c) Deposition kinetics analysis highlighting the acceleration role of LiF. (d) Stripping kinetics analysis highlighting the facilitation role of LiCOOH. (e) XRD patterns of Zn deposits verifying the preferred (002) orientation induced by the synergistic additives. (f) Radar chart summary of comprehensive performance metrics (kinetics, stability, reversibility).

**2.4 Electrochromic Performance and Kinetic Analysis of the Zn-RMED**

Switching speed is a critical performance metric for dynamic windows. While conventional Zn-based RMEDs typically suffer from sluggish kinetics,[4a, 5b, 13b, 14], the incorporation of LiF in our system has successfully bridged this gap, enabling rapid optical modulation. **Figure 3a** evaluates the deposition and stripping kinetics across various applied voltages. As expected, reaction rates increase with higher overpotentials; however, excessively negative potentials risk triggering the parasitic hydrogen evolution reaction (HER). Therefore, -0.7 V and +0.8 V were selected as the optimal working potentials to balance kinetic efficiency with electrochemical stability. Under these optimized conditions, the device demonstrates rapid switching times of 5.5 s (coloration) and 7.1 s (bleaching) (Figure 3b).

Figure 3d details the time-dependent deposition kinetics at −0.7 V, with corresponding optical spectra provided in Figure 3c. The device achieves a deep dark state (<0.1%) within 30 s. Furthermore, excellent coloration uniformity was confirmed by near-identical switching rates between central and edge regions (Figure 3e), a feature further validated by Figure S3 and Video S1. Optical bi-stability (or open-circuit memory), defined as the duration the dark state is maintained without power, is directly linked to energy-saving efficiency. As shown in Figure S4, a 30 s deposition at −0.7 V yields a memory time of only 300 seconds before spontaneous zinc dissolution restores transparency. Extending the deposition time to 200 s creates a thicker zinc layer, significantly prolonged retention to >30 minutes. Full recovery from this state requires 1 hour (for 30 s deposition) and 72 hours (for 200 s deposition), respectively, allowing for flexible adaptation to application-specific requirements. To further enhance this memory effect, benzotriazole (BTA) was explored as a corrosion inhibitor.[15] The addition of 200 ppm BTA significantly extends the optical retention: a 200-second deposition maintains the deep dark state for over 1 hour, with transmittance remaining below 20% even after 17 hours (Figure S5). Although this comes at a slight cost to switching speed, it offers a viable trade-off for applications requiring long-term state retention.

The electrochemical reversibility of the system was quantified by the Coulombic efficiency (CE). As shown in Figure 3f, the optimized device achieves a high CE of 99.3 %. near-unity value indicates highly reversible metal plating/stripping processes and the effective suppression of side reactions (e.g., electrolyte decomposition), which are crucial for long-term stability. In sharp contrast, control devices lacking LiCOOH or LiF exhibited significantly lower CEs of 92.9 % and 66.1 %, respectively (Figure S6), further confirming the necessity of the synergistic electrolyte design.

Finally, we evaluated the Coloration Efficiency, defined as the maximum optical density change per unit charge injected per unit charge density. While conventional

Zn-RMEDs (and even Cu-Zn systems)[16] often exhibit limited optical gain, our device achieves a high value of 29.33 cm² C⁻¹ (Figure 3g), implying that minimal charge injection is required to attain significant optical modulation. To assess robustness after deep cycling, Figure 3h examines the recovery following 300 s of deposition. The application of +0.8 V restored full transparency within 40 s, confirming exceptional reversibility. The cyclic voltammetry (CV) response (Figure 3i) further corroborates the stable redox behavior, distinguishing our optimized system from the inferior kinetics observed in additive-free controls (Figure S7).

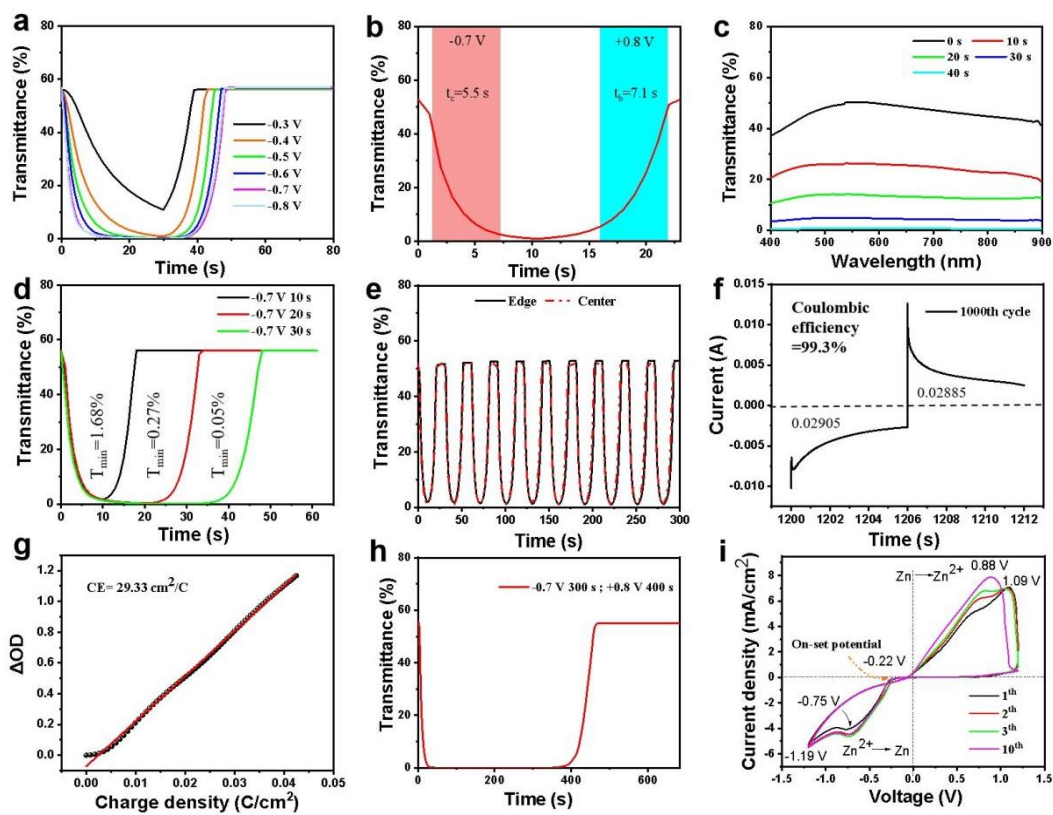

**Figure 3 Electrochromic performance and kinetic evaluation of the Zn-RMED.** (a) Voltage-dependent switching kinetics and current response analysis. (b) Dynamic transmittance modulation profile showing rapid response times at optimal potentials. (c) Evolution of Vis-NIR transmission spectra during the Zn deposition process. (d)

Transmittance decay kinetics as a function of deposition duration. (e) Spatiotemporal uniformity verification comparing switching profiles at the device center and edge. (f) Electrochemical Coulombic efficiency (CE) stability over continuous cycling. (g) Coloration efficiency plot. (h) Deep-cycling recovery test demonstrating full transparency restoration after prolonged (300 s) deposition. (i) Cyclic voltammetry (CV) characterization of the optimized redox behavior..

**2.5 Reflectance Characteristics and Mirror State Modulation**

The reflectivity of reversible metal electrodeposition devices (RMEDs) is governed by both the intrinsic dielectric properties of the metal and the morphological uniformity. Ag-RMEDs typically exhibit superior optical performance due to the high electrical conductivity of Ag. Though reported values vary widely between (17 %-89 %) depending on deposition quality[4d], Ag-RMEDs can achieve near-unity mirror states (>99 %)[17]. In contrast, Bi-Cu-RMEDs consistently maintain reflectivity above 50 %[18] and 40 %[19]. Zinc, intrinsically, exhibits lower specular reflectivity, with literature values typically ranging between 28 % and 43 %[4a]. However, through our synergistic interface engineering, we demonstrate that Zn-RMEDs can overcome this intrinsic limitation.

**Figure 4a** presents the reflectivity evolution over the initial cycles. The first cycle serves as an electrochemical activation step, after which the device enters a stable switching regime. The reflectivity modulates from 10 % to 60 % ($\Delta R$ = 50 %) over subsequent cycles. Notably, the time required to reach this high-reflectivity state ($t_{dep.}$=14.5 s) is longer than that for transmittance switching. This is because achieving a highly specular mirror requires the formation of a physically continuous and dense metal layer, rather than just a light-blocking particle network. We further investigated the voltage dependence. While increasing the overpotential beyond -0.7 V can marginally enhance reflectivity, it concurrently exacerbates the parasitic hydrogen evolution reaction (HER), leading to surface roughening. Thus, -0.7 V was selected as the optimal compromise between optical quality and electrochemical stability. A unique feature of our device is its asymmetric optical response (Figure 4c). When observed from the "Outdoor mode" (light incident from the ITO/glass substrate side), the device exposes the smooth, substrate-templated Zn interface, yielding a high reflectivity of >60 % that effectively rejects solar heat (Figure S8). Conversely, in "Indoor mode" (light incident from the electrolyte side), the rougher growth front of the Zn film scatters light, resulting in a lower matte-like reflectivity (~30 %). This design dual-mode design is advantageous, providing high solar rejection externally while maintaining a glare-free aesthetic internally.

To evaluate operational stability, reflectivity retention was characterized under open-circuit conditions (Figure 4d). Following a 30 s deposition, the reflectivity decayed from 60 % to baseline within 3 hours. This decay is faster than the transmittance loss because specular reflectivity is highly sensitive to surface microscopic roughening caused by spontaneous oxidative dissolution, even before significant mass loss occurs. To mitigate this, a minimal "maintenance bias" of -0.1 V was applied (Figure 4e). This cathodic potential effectively suppresses dissolution without triggering HER, extending the high-reflectivity state (>40 %) for over 168 hours. Upon returning to open-circuit conditions, the device retained 40 % reflectivity for 7 days, representing a 56-fold enhancement in optical memory compared to the unbiased state.

To further push the reflectivity limit, we introduced benzotriazole (BTA) as a crystallographic modifier (brightener). As shown in Figure S9, the addition of 200 ppm BTA dramatically boosted the peak reflectance to >80 %, surpassing the typical limits of Zn-based systems (Table S2). Furthermore, BTA adsorption forms a protective film that retards corrosion, under the same deposition conditions (-0.7 V, 30 s), the BTA-modified device maintained >20 % reflectance even after 18 hours of open-circuit rest. Finally, spectral chromaticity analysis (Figure 4f and Figure S10) confirms that color stability is voltage-mode dependent. While pulsed protocols induce coordinate shifts, constant-potential operation (-0.7 V), high-voltage driving, and stepped schemes exhibit negligible chromaticity variation, ensuring consistent visual aesthetics.

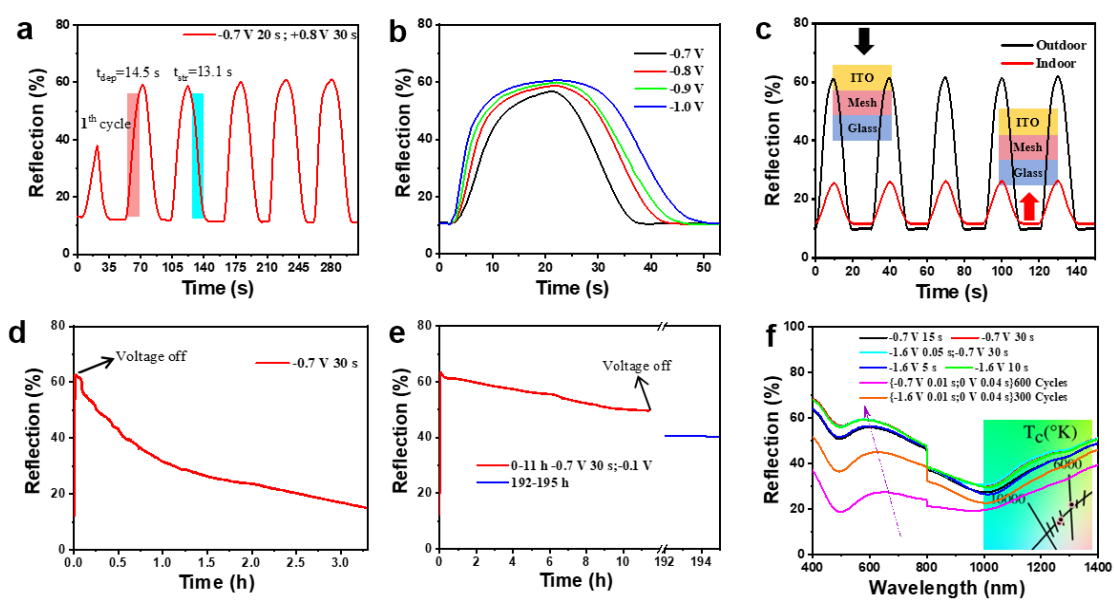

**Figure 4 Reflectance Characteristics and Mirror State Modulation:** (a) Dynamic

reflectivity evolution during the initial activation cycles. (b) Voltage-dependent reflectivity profiles indicating the optimal potential window. (c) Asymmetric optical response demonstrating dual-mode operation: high-reflection "Outdoor Mode" (light incident from ITO side) vs. low-reflection "Indoor Mode" (light incident from electrolyte side). (d) Reflectivity decay kinetics under open-circuit conditions (spontaneous self-discharge).

(e) Long-term optical retention test (>190 h). (f) Vis-NIR reflection spectra and corresponding CIE 1931 chromaticity evolution under various voltage.

## 2.6 Counter Electrode Engineering and Cycling Stability Mechanism

The design of the counter electrode (CE), acting as the reversible $Zn^{2+}$ reservoir, is paramount for stable device operation. Initially, zinc foil was selected due to its cost-effectiveness. However, as shown in Figure S11a, the foil electrode suffered from kinetic limitations, exhibiting decreased deposition and stripping rates. Non-uniform electric fields at the foil edges led to heterogeneous deposition and stripping. The geometric singularity leads to heterogeneous deposition, resulting in eventual delamination and irreversible adhesion to the ITO working electrode, thereby causing device failure (Figure S11b). Simulation of the potential distribution (Figure S11c) reveals that the "edge effect" causes severe electric field concentration at the foil periphery.

To circumvent these geometric constraints, a stainless-steel mesh was adopted as a mechanically robust scaffold. However, the pristine mesh lacks an intrinsic $Zn^{2+}$ source (**Figure 5a**). Pre-plating zinc in a standard electrolyte ($ZnSO_4$/$Na_2SO_4$/$H_3BO_3$) resulted in a disordered, dendritic morphology that compromised the initial optical transparency. To modulate the crystallization kinetics, glucose was introduced into the plating bath as a leveling agent. The glucose molecules preferentially adsorb onto active growth sites, effectively suppressing dendritic protrusion and inducing the formation of a dense, uniform zinc coating. Figure 5b presents the finite element simulation of the electric field modulus. The smoothed surface (representative of glucose modification) yields a uniform electric field distribution, contrasting sharply with the localized field hotspots observed on rough surfaces, thereby theoretically validating the dendrite-suppression mechanism.

Cycling stability is a paramount metric for the practical viability of dynamic windows. Assuming 1–2 daily switching events over a 20-year service life, a durability of >14,000 cycles is imperative. While Bi-Cu-RMED devices typically range from 200 to 2,500 cycles[5c, 18, 20] and Ag-RMED devices have achieved 2,000–5,000 cycles[6, 11, 21] achieving , achieving ultra-long-term stability in aqueous Zn-based systems remains a formidable challenge due to hydrogen evolution (HER), passivation by-products, and dendrite-induced short circuits. Unlike prior strategies that rely on complex "voltage compensation" protocols[4a], or "resting intervals"[22], to artificially extend lifespan, this work evaluates stability under rigorous constant voltage/time conditions.

Figure 5c presents the long-term transmittance cycling performance. Since the opaque mesh obstructs light path, measurements were normalized to the mesh transmittance (initial~ 80 %). Devices with unmodified electrodes failed rapidly within 3,500 cycles (Figure 5c, upper). In sharp contrast, the glucose-modified device sustained over 15,000 cycles at -0.7 V (5 s) / 0.8 V (8 s), with the optical modulation (ΔT) gradually decaying from 54 % to 37 % (Figure 5c, lower).

We conducted a comprehensive failure analysis to understand the decay mechanism. The reduction in maximum transmittance is attributed to the uneven electric field distribution caused by non-uniform film thickness, which slows dissolution kinetics. Conversely, the rise in minimum transmittance correlates with "dead Zn" accumulation and passivation. Furthermore, Figure S12 confirms the accumulation of HER-induced gas bubbles. Crucially, post-mortem analysis revealed that zinc deposition became confined to localized regions (Video S2). Conductivity mapping of the cycled ITO (Figure S13) indicated a near-total loss of conductivity in the "inactive" colored regions, confirming that electrochemical etching of the ITO substrate is a primary failure mode. This degradation can be potentially mitigated by protecting the ITO with a chemically inert layer (e.g., Pt). Additionally, the zinc mesh CE itself undergoes degradation; microscopic inspection (Figure S14) reveals a loss of metallic luster after cycling, indicative of irreversible zinc consumption.

Reflectance stability faces even stricter morphological requirements than transmittance. Figure S15 tracks the morphological evolution: while the film remains dense after 10 cycles, surface roughening initiates by 1,000 cycles, and severe disorder/darkening appears by 8,000 cycles. Since specular reflectivity is highly sensitive to roughness, the reflectance modulation (ΔR) degrades faster than ΔT. Nevertheless, the glucose-modified CE significantly outperforms the unmodified control (Figure 5d). While the control device failed to sustain reflectivity beyond 1,000 cycles, the modified device maintained robust switching for over 2,000 cycles with ΔR retaining ~50 % of its initial value. To benchmark this breakthrough, Table S2 provides a comprehensive comparison with 20 state-of-the-art electrochromic systems, highlighting our device's leading position in terms of cycling longevity and optical contrast.

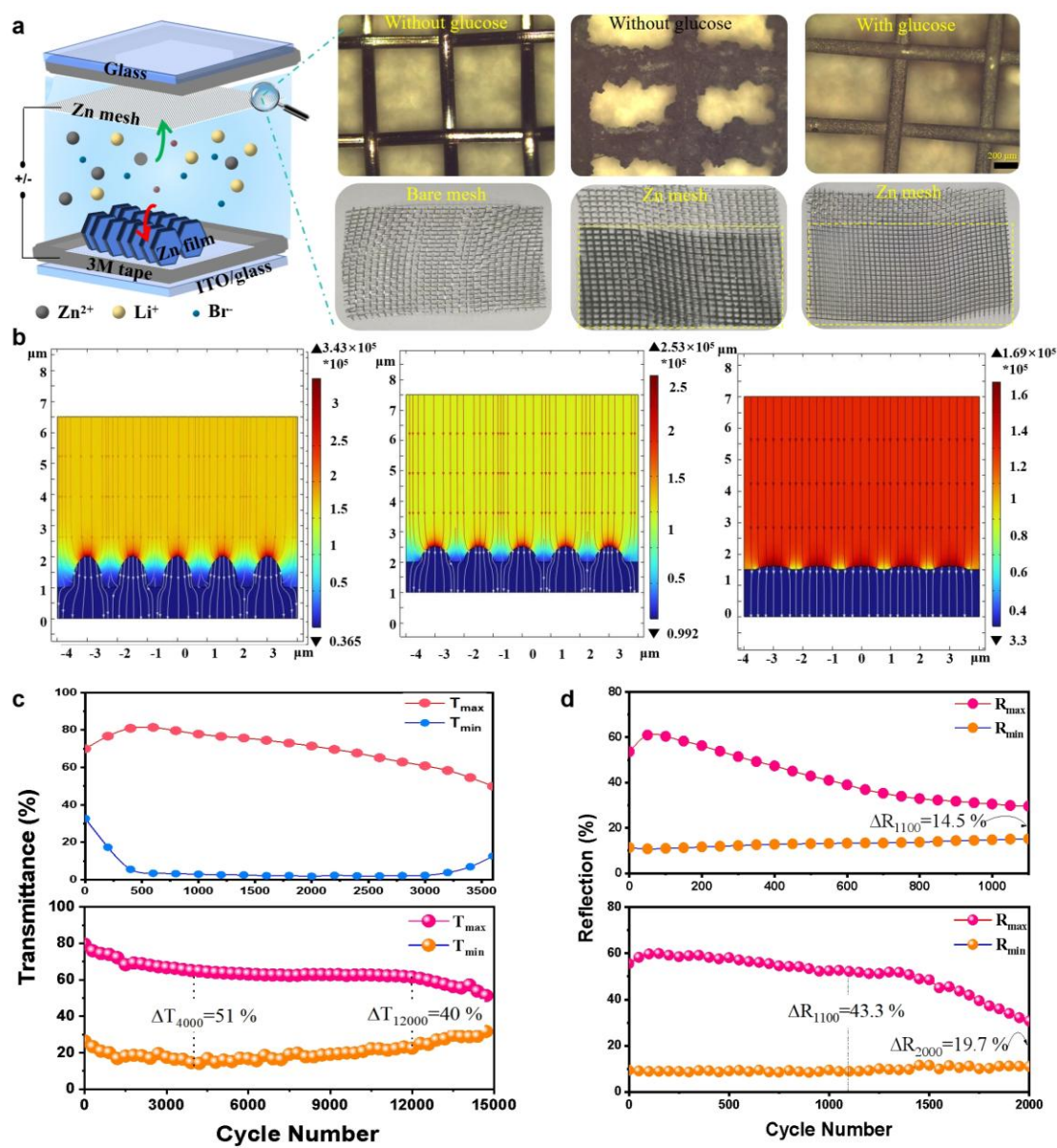

**Figure 5 Counter electrode optimization and cycling stability analysis.** (a) Schematic of the CE fabrication process: converting pristine stainless-steel mesh into a functional Zn reservoir via glucose-assisted electroplating. (b) Finite element simulation of the electric field distribution, demonstrating field homogenization on the glucose-modified smooth surface compared to the dendritic control. (c) Long-term transmittance cycling stability (at 600 nm) comparing the rapid failure of the unmodified device (upper) vs. the ultra-long lifespan (>15,000 cycles) of the glucose-modified device (lower). (d) Reflectance cycling stability comparison, highlighting the enhanced durability (>2,000 cycles) enabled by the modified counter electrode.

## 2.7 Environmental Adaptability: Broad-Temperature Operation and Thermal Management

The Zn-RMED is designed for adaptive climate control, dynamically regulating solar heat gain to match seasonal demands. In high-temperature environments (e.g., summer), the device switches to the Cooling Mode (mirror state), minimizing solar heat intake through enhanced reflection and attenuated Vis-NIR transmission. Conversely, in low-temperature conditions (e.g., winter), it transitions to the Heating Mode (transparent state) to maximize passive solar thermal harvesting (**Figure 6a**). To validate robustness under extreme heat, the device was subjected to continuous cycling at 80 °C on a hot plate (Figure S16a). Remarkably, it sustained over 3,000 cycles with minimal performance degradation (Figure S16b), verifying its suitability for operation in hot climates.

The thermal barrier performance was quantified via in situ infrared (IR) thermography (Figure S17). When subjected to a constant heat source (81.4 °C), the device in the mirror state maintained a significantly lower surface temperature (53.7 °C) compared to the heat source itself. This substantial temperature gradient (ΔT 30 °C) confirms the device's efficacy in blocking thermal radiation. Spectroscopic analysis (Figure 6b) further corroborates this: in the mirror state, transmittance remains suppressed below 1 % across the entire solar spectrum (400–2500 nm), demonstrating near-perfect shielding of visible and NIR light.

To assess stability under simultaneous thermal and optical stress, we conducted an in-situ transmittance monitoring test at 60 °C (Figure 6c). A continuous laser beam monitored the optical evolution of the device mounted on a vertical heater. After 5 hours of continuous operation (~1,636 cycles), the device retained 64 % of its initial transmittance modulation. While some degradation occurred, this result highlights robust resilience against thermally induced fatigue. Furthermore, a simulated "model house" experiment (Figure 6d, Figure S18) using a thermally insulated chamber under direct sunlight confirmed that the device in the colored state effectively lowered the internal temperature (measured by thermocouples TC2/TC3) compared to the transparent state, validating its practical heat insulation capacity.

A critical limitation of conventional DMSO-based electrolytes is their high freezing point (~18.55 ºC), which leads to solidification and functional failure in winter conditions. As shown in Figure S19, the standard electrolyte solidifies completely at -20 °C, rendering the device inoperable (though reversible upon thawing, see Video S3). To overcome this, we employed a binary solvent engineering strategy by introducing acetonitrile (ACN) as a cosolvent. Based on the principle of freezing point depression (colligative properties), the addition of 30 vol.% ACN disrupts the ordered molecular packing of DMSO, maintaining the liquid state even after 24 hours at -20 °C (Figure S20).

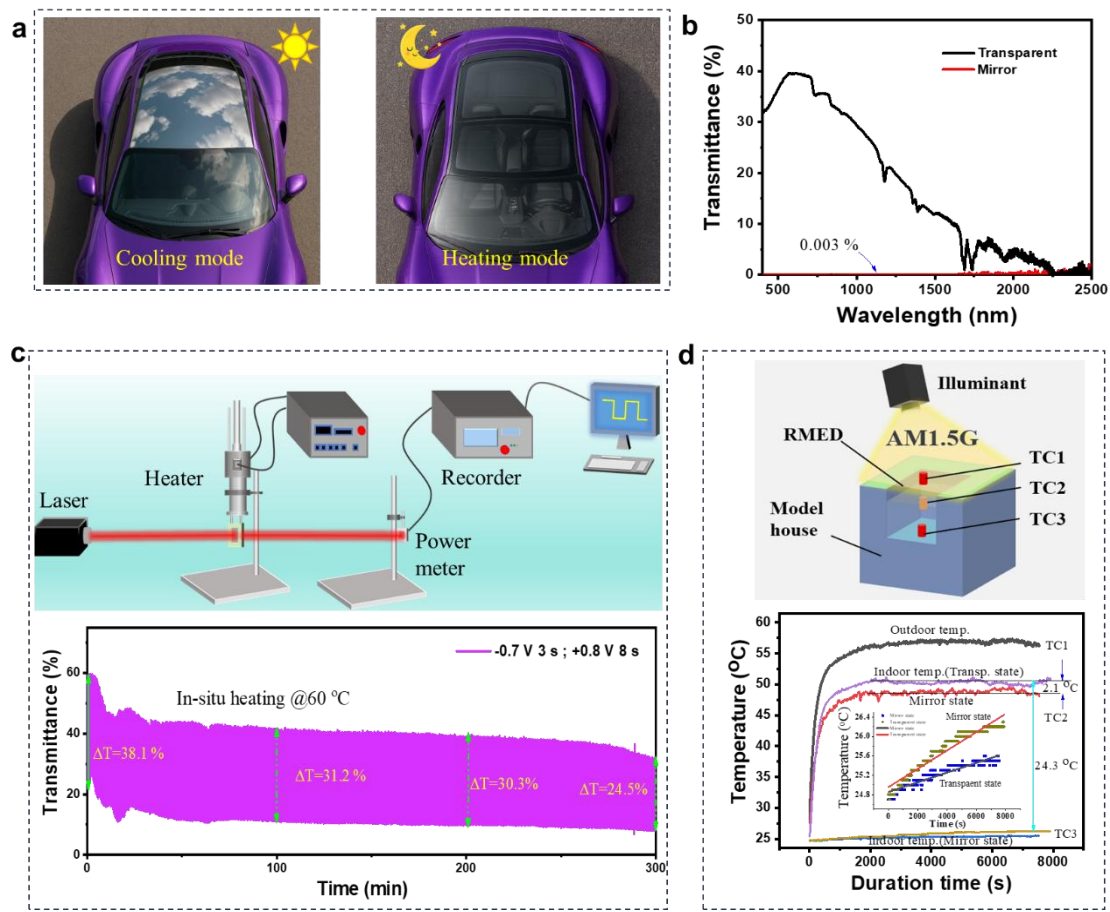

**Figure 6** All-climate thermal management and broad-temperature adaptability.

(a) Concept of adaptive climate control: switching between solar blocking (Summer/Cooling) and solar harvesting (Winter/Heating). (b) Vis-NIR spectra in the mirror state demonstrating <1 % transmittance across the solar band (400-2500 nm). (c) In situ monitoring of optical transmittance under continuous thermal stress at 60 °C. (d) Photothermal performance evaluation using a simulated model house, verifying the cooling effect under direct sunlight.

Leveraging this low-temperature electrolyte, the optimized Zn-RMED

demonstrated exceptional operability in freezing environments. As shown in Figure S21a and Video S4, the device achieved over 7,500 in-situ cycles at -20 °C, a record-high value for aqueous-based RMEDs. Post-cycling analysis (Figure S21b) revealed minor accumulation of insoluble precipitates and dendrites on the ITO surface, identifying the limiting factors for further lifespan extension. Nevertheless, these results establish the Zn-RMED as a viable technology for all-season thermal management.

## 3. Conclusions

In summary, we have developed a robust, high-contrast, and recyclable zinc-based dynamic window by implementing a synergistic electrolyte and interfacial engineering strategy. Our molecular dynamics (MD) simulations uncover a pivotal "dual-ion cooperative mechanism" that fundamentally governs the device's reversibility: strongly interacting F- anions (-200 kJ mol-1 dominate the inner solvation shell to guide uniform Zn deposition, while interfacial formate anions (HCOO- actively facilitate smooth stripping via proton-assisted dissolution. We demonstrate that this electrochemical synergy, combined with steric regulation from PVA and a glucose-modified counter electrode, effectively eliminates parasitic side reactions and dendrite growth.

Driven by this mechanism, our optimized Zn-RMED delivers state-of-the-art performance, including high Coulombic efficiency (>99.0 %), superior reflectivity (>60 %), and rapid switching kinetics ($t_c$=5.5 s/$t_b$=7.1 s). Most notably, the device achieves a record-breaking lifespan of over 15,000 cycles in transmittance mode and 2,000 cycles in reflectance mode (Δ R > 50 %). Furthermore, we validate the device's "all-climate" adaptability: it sustains stable operation at elevated temperatures (80 °C) for over 3,000 cycles and, with our binary solvent strategy, performs robustly at -20 °C for over 7,500 cycles. This work establishes a new performance benchmark for aqueous optoelectronics and provides a scalable, atomistic-level design protocol for next-generation energy-saving smart windows.

## 4. Experimental Section
### 4.1 Preparation of Electrolyte and Fabrication of Electrodes

The electrochromic electrolyte was prepared by dissolving 0.3 M $ZnBr_2$ (99.9 %), 0.4 M LiCOOH (99.9 %), 0.2 M LiF (99.9 %), and 30 mg poly(vinyl alcohol) (PVA, Mw ≈ 30,000~70,000) in 5 ml anhydrous DMSO. The mixture was stirred at 60 °C for two hours until a transparent homogeneous solution was obtained. For the working electrode, Indium Tin Oxide (ITO) glass substrates were ultrasonically cleaned sequentially in deionized water, acetone, and isopropanol for 10 minutes each, followed by drying under a nitrogen stream. For the counter electrode, a commercial stainless-steel mesh (30 mesh) was utilized as the scaffold. Prior to device assembly, the mesh underwent zinc electroplating to form a reversible ion reservoir. The electroplating electrolyte consisted of 0.5 M $ZnSO_4 \cdot 7H_2O$, 1.0 M $Na_2SO_4$, and 0.4 M $H_3BO_3$. To induce uniform zinc deposition and suppress dendrite growth, glucose was introduced into the electrolyte as a leveling agent. The electroplating was performed using a two-electrode configuration with a zinc foil as the counter electrode and the stainless-steel mesh as the working electrode (experimental setup shown in Figure S22).

**4.2 Characterization and measurements**

The in-situ transmittance measurements were conducted using a UV-Vis-NIR spectrophotometer (UV-2600, Shimadzu) coupled with an electrochemical workstation (CHI 660E, Chenhua). Reflectance spectra were calibrated using a standard aluminum mirror (STD-M, Wenyi Photoelectric Technology) as the 100 % reflectance reference. The surface morphology of the electrodes was characterized by Scanning Electron Microscopy (SEM, Phenom XL, FEI, Netherlands). For thermal performance evaluation, a high-temperature heating stage (JW-400XT, Wuhan Junwei Technology) was employed for heating experiments. A solar simulator light source (LCS-100, Newport) and a thermocouple (TA612C) were used for photothermal testing. Low-temperature performance was evaluated using a liquid nitrogen cryostat (C16070080, East Changing) equipped with a precise temperature controller (TC 202, East Changing). The light intensity for the optical modulation tests was measured using an optical power meter (PM 100D, Thorlabs) with a 660 nm laser diode (Shenzhen LEO Optoelectronics) as the light source.

**4.3 Molecular dynamics (MD) calculation**

Molecular dynamics (MD) simulations were employed with GROMACS software[23]. A time step of 1 fs was selected, with a cutoff distance of 1.0 nm for both the van der Waals interactions and electrostatic interactions. The electrostatic interactions were calculated by the particle mesh Ewald method (PME). The nonbonded interaction contains van deer Waals (vdW) and electrostatic interaction. The box of the system is 5.75×5.75×5.75 $nm^3$. The system contains 50 PVA chains, 1400 DMSO, 30 $Zn^{2+}$, 60 $Br^-$, 60 $Li^+$, 40 $HCOO^-$ and 20 $F^-$. The interaction parameters were taken from GAFF force field and RESP atomic charge. The simulation was firstly energy minimized, and then NPT ensemble was conducted at 298.15 K and 1 bar. Temperature and pressure were regulated using the velocity-rescale thermostat and Berendsen barostat, respectively. For the production simulations, the simulations were carried out for 30 ns to ensure the system reaching equilibrium, and the last 2 ns of the equilibrium state were used for subsequent post-processing analysis.

**Author contributions**

**Fei Peng, Zhoutao Yang:** Methodology, Investigation, Formal analysis, Data curation，Conceptualization. **Zhichao Liang, Yijia Chen:** Methodology, Simulation, Formal analysis, Data curation. **Wenjie Mai:** Supervision, Funding acquisition. **Chuanxi Zhao:** Writing – review & editing, Supervision, Funding acquisition, Data curation, Conceptualization.

**Competing Interests statement**

The authors declare no competing interests.

Correspondence and requests for materials should be addressed to C. Z. or W. M.

**Acknowledgments**

This work was financially supported by National Natural Science Foundation of China (52172202); Guangdong Basic and Applied Basic Research Foundation (2024A1515011879, 2022A1515010049, 2023A1515011994, 2023A1515030163), Science and Technology Planning Project of Guangzhou, China (Grant Nos. 202102080229, 201605030008).